\documentclass[aps,pre,reprint,superscriptaddress,showpacs,preprintnumbers,amsmath,amssymb]{revtex4-2}
\usepackage{graphicx}
\graphicspath{{fig/}}
\usepackage{dcolumn}
\usepackage{bm}
\begin{document}
\preprint{Preprint}
\title{Relaxation dynamics and long-time tails explain shear-induced diffusion of soft athermal particles near jamming}
\author{Kuniyasu Saitoh}
\affiliation{Department of Physics, Faculty of Science, Kyoto Sangyo University, Kyoto 603-8555, Japan}
%
\author{Takeshi Kawasaki}
\affiliation{D3 Center,  The University of Osaka, Toyonaka, Osaka 560-0043, Japan}
\affiliation{Department of Physics, The University of Osaka, Toyonaka, Osaka 560-0043, Japan}
\date{\today}
\begin{abstract}
We numerically study shear-induced diffusion of soft athermal particles in two dimensions.
The Green-Kubo (GK) formula is applicable to the shear-induced diffusion coefficient,
where both mean squared transverse velocity and relaxation time included in the GK formula are well described by critical scaling near jamming.
We show that the auto-correlation function of transverse velocities is stretched exponential if the system is below jamming or shear rate is large enough.
However, if the system is above jamming and the shear rate is sufficiently small,
the auto-correlation function exhibits a long-time tail such that time integral in the GK formula diverges in two dimensions.
We propose an empirical scaling relation for the critical exponents and show that the long-time tail is consistent with the divergence of the shear-induced diffusion coefficient.
\end{abstract}
\maketitle
%
Soft athermal particles such as foam, emulsions, colloidal suspensions, and granular materials are ubiquitous,
and a better understanding of their transport properties is of central importance to technology,\
e.g.\ for manufacturing of food, pharmaceutical and personal care products \cite{bird,review-rheol0}.
It has been well recognized that soft athermal particles exhibit a \emph{jamming transition} at critical packing fraction $\phi_J$ \cite{gn3,gn4,gn1},
where their mechanical \cite{gn1,vm4} and rheological properties \cite{rheol0,pdf1,rheol11,rheol15,rl0,saitoh15} can be understood in terms of critical phenomena.
For instance, flow curves are nicely explained by critical scaling:
both viscosity divergence \cite{muI3,rheol8,rheol10,vm-unjam4,Suzuki_Hayakawa} and scaling of yield stress \cite{rheol0,rheol5,pdf1}
are described by power-laws of the proximity to jamming, $\Delta\phi\equiv\phi-\phi_J$, where $\phi$ is the packing fraction of the particles.

In contrast, \emph{shear-induced diffusion}, which is relevant to mixing and segregation of soft athermal particles \cite{diff_shear_md0},
has been studied with the focus on its dependence on a shear rate.
Using natural length and time scales,\ i.e.\ the particle diameter $d_0$ and inverse of shear rate $\dot{\gamma}^{-1}$,
one can expect the scaling of shear-induced diffusion coefficient as $D\sim d_0^2\dot{\gamma}$ \cite{diff_shear_exp1,diff_shear_exp2,diff_shear_exp3,diff_shear_exp4,diff_shear_md11}.
However, increasing the shear rate, researchers observed a crossover from the linear scaling to sub-linear scaling,\
i.e.\ from $D\propto\dot{\gamma}$ to $\dot{\gamma}^{0.8}$ \cite{diff_shear_exp5,diff_shear_md1,diff_shear_md2,dh_md2}.
Another sub-linear scaling, $D\propto\dot{\gamma}^{0.78}$, was also found at the onset of jamming transition, $\phi\simeq\phi_J$ \cite{diff_shear_md3}.
Moreover, the scaling of shear-induced diffusion coefficient was modified as $D\sim d_0^2\dot{\gamma}/\sqrt{I}$ with the inertia number $I$
in order to take account of the influence of collective motions of the particles \cite{diff_shear_md4,diff_shear_md5,diff_shear_md6}.

Recently, we demonstrated a scaling data collapse of the shear-induced diffusion coefficient near jamming \cite{saitoh14},
predicting not only the crossover but also the critical divergence of $D$ (that is consistent with the inverse proportionality with the pressure $p$,\ i.e.\ $D\propto p^{-1}$ \cite{diff_shear_exp4}).
It was shown that the diffusion coefficient scales as $D\sim d_0\xi\dot{\gamma}$ with a characteristic size of collective motions, $\xi$ \cite{diff_shear_md4,dh_qs1}.
If the system is above jamming ($\phi>\phi_J$) and sheared at a sufficiently small shear rate,
the size of collective motions extends to the system length $L$,\ i.e.\ $\xi\sim L$,
so that the diffusion coefficient over the shear rate is linear in the system length as $D/\dot{\gamma}\propto L$ \cite{saitoh14}.
This agrees with the previous studies of sheared amorphous solids \cite{diff_shear_md7,diff_shear_md8,diff_shear_md9,diff_shear_md10}
though the finite-size effect on $D/\dot{\gamma}$ was explained as a result of spatially correlated flip events.
In recent years, (slightly) different finite-size scaling, $D/\dot{\gamma}\propto L^{1.05}$ \cite{diff_shear_ep1,diff_shear_ep2} and $L^{1.5}$ \cite{diff_shear_ep0},
was also reported by numerical simulations of elastoplastic model.

The previous studies have revealed the relationship between the shear-induced diffusion coefficient and characteristic length scales in the system.
However, transport properties of thermal systems such as classical fluids and glasses are explained by relaxation dynamics \cite{Hansen,SE_break0,SE_break1},
and little is known about the role of relaxation dynamics in the shear-induced diffusion of soft athermal particles.

In this Letter, we investigate the shear-induced diffusion of soft athermal particles by molecular dynamics (MD) simulations.
Employing the Green-Kubo (GK) relation for the diffusivity,
we clarify the link between diffusion coefficient $D$ and relaxation time extracted from the velocity auto-correlation function of the particles.
We show that (i) mean squared particle velocities obey critical scaling near jamming,
(ii) the relaxation time is also explained by critical scaling only if the system is not in a quasi-static regime above jamming,
and (iii) \emph{long-time tails} are observed in the quasi-static regime above jamming such that $D$ formulated by the GK relation diverges in the thermodynamic limit, $L\rightarrow\infty$.
We use the reduced GK formula to derive (iv) an empirical scaling relation for the critical exponents.
In contrast to the previous works, where $D$ is connected to the length scales \cite{diff_shear_md4,diff_shear_md5,diff_shear_md6,saitoh14}
and finite-size effects on $D$ are interpreted as the system spanning flip events \cite{diff_shear_md7,diff_shear_md8,diff_shear_md9,diff_shear_md10,diff_shear_ep1,diff_shear_ep2,diff_shear_ep0},
our analysis is based on the time scale, where the divergence of $D$ is explained by the long-time tail.

\emph{Numerical methods.}---
We perform MD simulations of soft athermal particles in two dimensions.
Our system consists of an equal number of small and large particles with diameters, $d_0$ and $1.4d_0$, respectively \cite{gn1}.
The total number of particles is $N=8192$ (we examine finite-size effects in later)
and their packing fraction $\phi$ is controlled around the jamming transition density, $\phi_J\simeq0.8433$ \cite{rheol0}.
A repulsive force between the particles, $i$ and $j$, in contact is modeled by elastic force, $\bm{f}_{ij}^\mathrm{e}=k\delta_{ij}\bm{n}_{ij}$,
where $k$ is the stiffness and $\bm{n}_{ij}\equiv\bm{r}_{ij}/|\bm{r}_{ij}|$ with the relative position $\bm{r}_{ij}\equiv\bm{r}_i-\bm{r}_j$ is the normal unit vector.
The elastic force is linear in the overlap $\delta_{ij}\equiv R_i+R_j-|\bm{r}_{ij}|>0$, where $R_i$ ($R_j$) is the radius of the particle $i$ ($j$).
We also add a damping force to every particle as $\bm{f}_i^\mathrm{d}=-\zeta\left\{\bm{v}_i-\bm{u}(\bm{r}_i)\right\}$,
where $\zeta$, $\bm{v}_i$, and $\bm{u}(\bm{r})$ are the damping coefficient, particle velocity, and external flow field, respectively.
Note that the stiffness and damping coefficient determine a microscopic time scale as $t_0\equiv\zeta/k$.
To simulate simple shear flows of the particles, we impose the external flow field in the $x$-direction as $\bm{u}(\bm{r})=(\dot{\gamma}y,0)$ under the Lees-Edwards boundary condition \cite{lees},
where $\dot{\gamma}$ is introduced as a shear rate.
In our simulations, motions of the particles are described by \emph{overdamped dynamics} \cite{rheol0,rheol7,pdf1},\ i.e.\ $\sum_{j\neq i}\bm{f}_{ij}^\mathrm{e}+\bm{f}_i^\mathrm{d}=\bm{0}$,
where we numerically integrate the particle velocity, $\bm{v}_i=\bm{u}(\bm{r}_i)+\zeta^{-1}\sum_{j\neq i}\bm{f}_{ij}^\mathrm{el}$, with a small time increment, $\Delta t = 0.1t_0$.

In the following, we only analyze the data in a steady state, where the shear strain applied to the system exceeds unity, and scale every time and length by $t_0$ and $d_0$, respectively.
In addition, we vary $\phi$ and $\dot{\gamma}$ in the ranges, $0.8\leq\phi\leq 0.9$ and $10^{-7}\leq\dot{\gamma}t_0\leq 10^{-2}$, respectively.

\emph{Diffusion coefficient and the Green-Kubo relation.}---
We quantify shear-induced diffusion of the particles by the transverse component of mean squared displacement (MSD) \cite{saitoh14},
\begin{equation}
\Delta(\tau)^2 = \left\langle \frac{1}{N}\sum_{i=1}^N \Delta y_i(t+\tau)^2 \right\rangle_t~.
\label{eq:MSD}
\end{equation}
Here, $\Delta y_i(t+\tau)\equiv y_i(t+\tau)-y_i(t)$ is the $y$-component of particle displacement for the duration $\tau$
and we take the ensemble average $\langle\dots\rangle_t$ by changing the initial time $t$ in a steady state.
The MSD is linear in $\tau$ if the amount of shear strain (for the duration $\tau$) exceeds unity,\ i.e.\ if $\gamma\equiv\dot{\gamma}\tau>1$.
Thus, a diffusion coefficient can be defined as $D=\lim_{\tau\rightarrow\infty}\Delta(\tau)^2/2\tau$.

In Ref.\ \cite{saitoh14}, we demonstrated a scaling data collapse of the diffusion coefficient near jamming
and numerically confirmed its critical scaling as $D\sim |\Delta\phi|^{-\nu}\dot{\gamma}$ ($\phi<\phi_J$), $D\sim |\Delta\phi|^{0.3\lambda-\nu}\dot{\gamma}^{0.7}$ ($\phi>\phi_J$),
and $D\sim \dot{\gamma}^{1-\nu/\lambda}$ ($\Delta\phi$-independent \emph{critical regime}).
As shown in the Supplemental Materials (SM) \cite{SM}, the critical exponents are quantitatively estimated as $\nu\simeq 0.939$ and $\lambda\simeq 3.87$
\footnote{The exponents were roughly estimated as $\nu=1.0$ and $\lambda=4.0$ in Ref.\ \cite{saitoh14}.}
which well agree with the previous simulation \cite{diff_shear_md3}.
Because the diffusion coefficient over the shear rate is proportional to the length scale as $D/\dot{\gamma}\propto\xi$ \cite{saitoh14,SM},
our result below jamming implies the scaling, $\xi\propto D/\dot{\gamma}\sim |\Delta\phi|^{-\nu}\simeq |\Delta\phi|^{-0.939}$,
which is consistent with the divergence of correlation length at the jamming transition,\ i.e.\ $\xi\sim |\Delta\phi|^{-1}$ \cite{nafsc2,dh_qs1,rheol17}.
In addition, our scaling indicates a crossover from $D\propto\dot{\gamma}$ to $\dot{\gamma}^{1-\nu/\lambda}\simeq\dot{\gamma}^{0.757}$ with the increase of shear rate,
which also agrees with the previous studies \cite{diff_shear_exp5,diff_shear_md1,diff_shear_md2,dh_md2,diff_shear_md3}.

On the other hand, the diffusion coefficient is formulated by the GK relation as \cite{Hansen,diff_shear_md3}
\begin{equation}
D = \langle v_y^2\rangle\int_0^\infty C(\tau) d\tau~.
\label{eq:GK}
\end{equation}
Here, $\langle v_y^2\rangle\equiv \langle N^{-1}\sum_{i=1}^N v_{iy}(t)^2\rangle_t$ is the $y$-component of mean squared particle velocity,
where $\langle\dots\rangle_t$ is the time average in a steady state.
On the right-hand-side of Eq.\ (\ref{eq:GK}), $C(\tau)=\langle v_y(t+\tau)v_y(t)\rangle_t/\langle v_y^2\rangle$ is the normalized velocity auto-correlation function.

\begin{figure}
\includegraphics[width=\columnwidth]{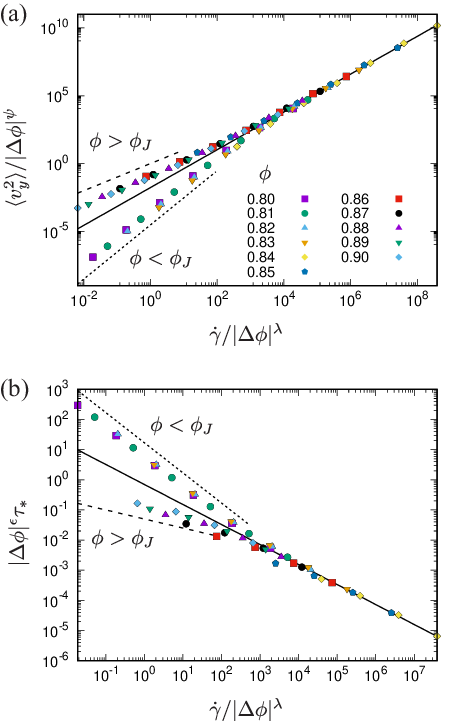}
\caption{
Scaling data collapses of (a) the mean squared transverse velocity $\langle v_y^2\rangle$ and (b) relaxation time $\tau_\ast$,
where the scaling exponents are given by $\lambda=3.87$, $\psi=5.39$, and $\epsilon=2.57$.
The solid lines have the slopes, (a) $\psi/\lambda$ and (b) $-\epsilon/\lambda$, and the symbols represent the packing fraction $\phi$ as listed in the legend of (a).
\label{fig:critical}}
\end{figure}

\emph{Critical scaling of the mean squared velocity.}---
As the diffusion coefficient $D$ exhibits critical behavior near jamming \cite{diff_shear_md3,saitoh14},
the mean squared transverse velocity $\langle v_y^2\rangle$ in the GK formula (Eq.\ \eqref{eq:GK}) is described by critical scaling.
Figure \ref{fig:critical}(a) displays a scaling data collapse of $\langle v_y^2\rangle$, where the packing fraction $\phi$ increases as listed in the legend.
In this figure, the scaling exponent is given by $\psi\simeq 5.39$ ($\lambda$ is the same with that for $D$).
The solid line has the slope $\psi/\lambda$, representing the $\Delta\phi$-independent critical regime.
In the SM \cite{SM}, we show how to estimate $\psi$ quantitatively and confirm that the $x$-component of mean squared velocity fluctuation,\
i.e.\ $\langle \delta v_x^2\rangle$ with $\delta v_x\equiv v_x-\dot{\gamma}y$, exhibits almost the same critical behavior.
This means that the critical scaling of \emph{effective temperature} \cite{rheol2,rheol3,rheol4,rheol5} of our system is similar to that of $\langle v_y^2\rangle$ (Fig.\ \ref{fig:critical}(a)).

In our MD simulations, the energy injected by shear is given by $\dot{\gamma}\sigma$ with $\sigma$ the shear stress of the system.
The injected energy is dissipated by the damping force $\bm{f}_i^\mathrm{d}$ such that the rate of energy dissipation $\chi$ is approximated as $\chi\sim\langle v_y^2\rangle$ \cite{CST0,CST1,pdf1}.
Therefore, due to the energy balance ($\dot{\gamma}\sigma\sim\chi$), the mean squared velocity is quadratic in the shear rate as $\langle v_y^2\rangle \sim \eta\dot{\gamma}^2$,
where $\eta\equiv\sigma/\dot{\gamma}$ is defined as shear viscosity.
The shear viscosity is constant ($\eta=\mathrm{const.}$) if the system is in the Newtonian regime,\ i.e.\
if the system is below jamming, $\phi<\phi_J$, and the shear rate is sufficiently small, $\dot{\gamma}t_0\ll 1$.
In contrast, if the system is above jamming, $\phi>\phi_J$, the shear stress exhibits yield stress $\sigma_Y$ in the quasi-static limit, $\dot{\gamma}\rightarrow 0$ \cite{review-rheol0}.
In this case, the energy balance equation tells us that the mean squared velocity is linear in the shear rate as $\langle v_y^2\rangle \sim \dot{\gamma}\sigma_Y$.

Taking account of the dependence of $\langle v_y^2\rangle$ on $\dot{\gamma}$, we describe our results in Fig.\ \ref{fig:critical}(a) as
\begin{equation}
\frac{\langle v_y^2\rangle}{|\Delta\phi|^\psi} = \mathcal{F}\left(\frac{\dot{\gamma}}{|\Delta\phi|^\lambda}\right)
\label{eq:scaling_vy2}
\end{equation}
with a scaling function $\mathcal{F}(x)$.
In the Newtonian regime, the scaling function is quadratic, $\mathcal{F}(x)\sim x^2$ (dotted line in Fig.\ \ref{fig:critical}(a)),
so that $\langle v_y^2\rangle \sim |\Delta\phi|^{\psi-2\lambda}\dot{\gamma}^2 \simeq |\Delta\phi|^{-2.35}\dot{\gamma}^2$.
Since $\langle v_y^2\rangle \sim \eta\dot{\gamma}^2$ in the Newtonian regime, this result indicates $\eta\sim|\Delta\phi|^{-2.35}$
which well agrees with the viscosity divergence near jamming \cite{rheol0,rheol11,rheol15}.
If the system is in the quasi-static regime above jamming, the scaling function is linear, $\mathcal{F}(x)\sim x$ (dashed line in Fig.\ \ref{fig:critical}(a)),
such that $\langle v_y^2\rangle \sim |\Delta\phi|^{\psi-\lambda}\dot{\gamma} \simeq |\Delta\phi|^{1.52}\dot{\gamma}$.
Because $\langle v_y^2\rangle \sim \dot{\gamma}\sigma_Y$ above jamming,
this result means that the yield stress scales as $\sigma_Y\sim|\Delta\phi|^{1.52}$ which is consistent with the previous result, $\sigma_Y\sim|\Delta\phi|^{3/2}$ \cite{rheol5,pdf1}.
In the $\Delta\phi$-independent critical regime, the scaling function is $\mathcal{F}(x)\sim x^z$ with the exponent $z$.
From Eq.\ \eqref{eq:scaling_vy2}, we find $\langle v_y^2\rangle \sim |\Delta\phi|^{\psi-\lambda z}\dot{\gamma}^z$ which should not depend on $\Delta\phi$.
Therefore, the exponents satisfy $\psi-\lambda z=0$,\ i.e.\ $z=\psi/\lambda$, as indicated by the solid line in Fig.\ \ref{fig:critical}(a).
%

\begin{figure}
\includegraphics[width=\columnwidth]{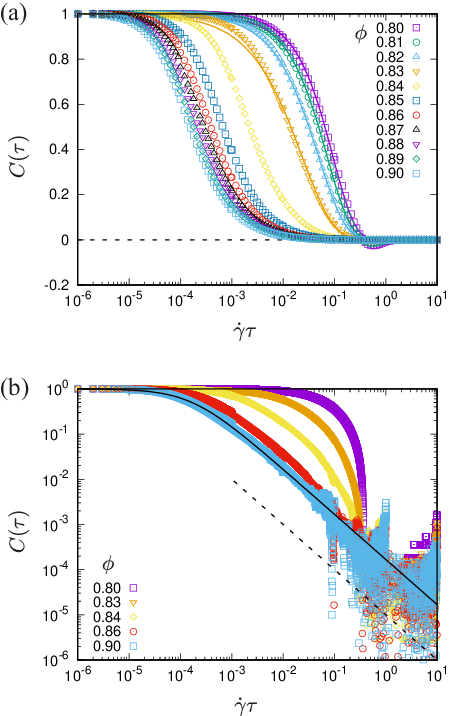}
\caption{
(a) Semi-logarithmic and (b) double-logarithmic plots of the normalized velocity auto-correlation function $C(\tau)$ (symbols),
where $\dot{\gamma}t_0=10^{-6}$ and $\phi$ varies as listed in the legends.
The horizontal axes are scaled by $\dot{\gamma}$ and the solid lines are the fits to $C(\tau)$ (see the text for the details).
The dashed line in (a) indicates zero and that in (b) shows the power-law decay, $C(\tau)\sim (\dot{\gamma}\tau)^{-1}$.
\label{fig:autocorrelation}}
\end{figure}

\emph{Critical scaling of relaxation time.}---
Next, we analyze the normalized velocity auto-correlation function $C(\tau)$ in the GK formula (Eq.\ \eqref{eq:GK}).
Figure \ref{fig:autocorrelation}(a) displays our numerical results of $C(\tau)$ (symbols),
where $\dot{\gamma}t_0=10^{-6}$ is used in MD simulations and $\phi$ varies as listed in the legend.
To describe our numerical results, we introduce a stretched exponential function as $C(\tau) = \exp[-(\tau/\tau_\ast)^\alpha]$,
where $\alpha$ and $\tau_\ast$ are the stretching exponent and relaxation time, respectively \cite{diff_shear_md1,diff_shear_md2}.
Adjusting the two parameters ($\alpha$ and $\tau_\ast$), we can see good agreements between the data below jamming, $\phi<\phi_J$, and stretched exponential functions (solid lines in Fig.\ \ref{fig:autocorrelation}(a)).
Note that $\alpha$ converges to unity if $\phi<\phi_J$ and $\dot{\gamma}\rightarrow 0$ so that $C(\tau)$ exponentially decays to zero in the Newtonian regime \cite{SM}.
We also find that the data above jamming, $\phi>\phi_J$, are well fitted to the stretched exponential function if the shear rate is large enough,\ i.e.\ $\dot{\gamma}t_0\geq 10^{-5}$.
However, $C(\tau)$ cannot be fitted to the stretched exponential function in a quasi-static regime above jamming,\ i.e.\ $\phi>\phi_J$ and $\dot{\gamma}t_0\leq 10^{-6}$.

The relaxation time $\tau_\ast$ extracted from $C(\tau)$ exhibits critical behavior near jamming.
Figure \ref{fig:critical}(b) shows a scaling data collapse of $\tau_\ast$, where the scaling exponent is estimated as $\epsilon=2.57$ (see the SM \cite{SM}).
The solid line with the slope $-\epsilon/\lambda$ represents the $\Delta\phi$-independent critical regime.
In this figure, $\tau_\ast$ cannot be defined for quasi-static flows above jamming ($\phi>\phi_J$ and $\dot{\gamma}t_0\leq 10^{-6}$).

To describe the data in Fig.\ \ref{fig:critical}(b), we introduce a scaling function $\mathcal{G}(x)$ as
\begin{equation}
|\Delta\phi|^\epsilon\tau_\ast = \mathcal{G}\left(\frac{\dot{\gamma}}{|\Delta\phi|^\lambda}\right)~.
\label{eq:scaling_tau}
\end{equation}
Based on our numerical results, we empirically determine the scaling function as follows.
In the Newtonian regime, the scaling function is $\mathcal{G}(x)\sim x^{-1}$ (dotted line in Fig.\ \ref{fig:critical}(b))
so that the relaxation time scales as $\tau_\ast\sim|\Delta\phi|^{\lambda-\epsilon}\dot{\gamma}^{-1}$.
On the other hand, if the system is above jamming, the scaling function is $\mathcal{G}(x)\sim x^{-0.3}$ (dashed line),
where the relaxation time scales as $\tau_\ast\sim|\Delta\phi|^{0.3\lambda-\epsilon}\dot{\gamma}^{-0.3}$.
In the $\Delta\phi$-independent critical regime, the scaling function is $\mathcal{G}(x)\sim x^w$ (solid line),
where the exponents satisfy $\lambda w + \epsilon=0$,\ i.e.\ $w=-\epsilon/\lambda\simeq 0.664$.
This result is reminiscent of the sub-linear scaling of relaxation time predicted by the nonequilibrium mode-coupling theory, $\tau_\mathrm{MCT}\sim\dot{\gamma}^{-0.8}$ \cite{MCT0}.
%

\begin{figure}
\includegraphics[width=\columnwidth]{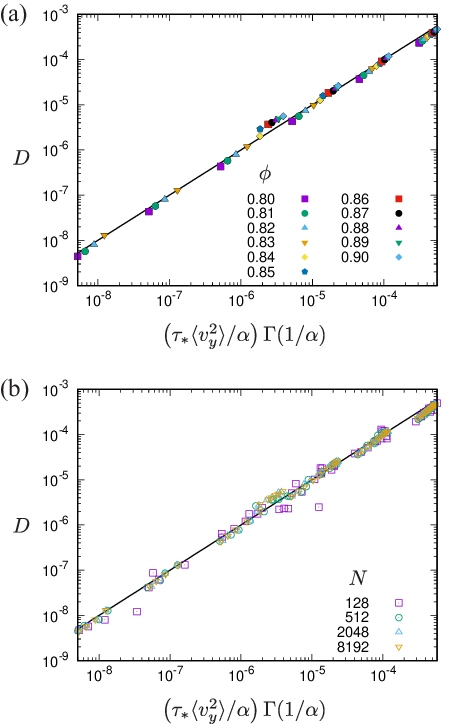}
\caption{
(a) A scatter plot of the diffusion coefficient $D$ extracted from the transverse MSD (Eq.\ \eqref{eq:MSD})
and the right-hand-side of Eq.\ \eqref{eq:GK-stretched},\ i.e.\ $(\tau_\ast\langle v_y^2\rangle/\alpha)\Gamma(1/\alpha)$.
The system size is $N=8192$ and the packing fraction $\phi$ increases as listed in the legend.
(b) System size dependence of the reduced GK formula (Eq.\ \eqref{eq:GK-stretched}), where the number of particles $N$ increases as listed in the legend.
In both (a) and (b), the solid lines represent $D=(\tau_\ast\langle v_y^2\rangle/\alpha)\Gamma(1/\alpha)$.
\label{fig:D-GK}}
\end{figure}

\emph{The reduced GK formula and a scaling relation.}---
We validate the GK relation and show that it is independent of the system size
as long as the system is \emph{not} in the quasi-static regime above jamming ($\phi>\phi_J$ and $\dot{\gamma}t_0\leq 10^{-6}$).
Because the velocity auto-correlation function is well described by the stretched exponential function,
we substitute $C(\tau) = \exp[-(\tau/\tau_\ast)^\alpha]$ into the GK formula.
Calculating the time integral on the right-hand-side of Eq.\ \eqref{eq:GK}, we find that the diffusion coefficient is given by
\begin{equation}
D = \frac{\tau_\ast\langle v_y^2\rangle}{\alpha}\Gamma\left(\frac{1}{\alpha}\right)~,
\label{eq:GK-stretched}
\end{equation}
where $\Gamma(x)$ is the gamma function \cite{SM}.
As shown in Fig.\ \ref{fig:D-GK}(a), $D$ extracted from the transverse MSD (Eq.\ \eqref{eq:MSD}) and the right-hand-side of Eq.\ \eqref{eq:GK-stretched} coincide.
In addition, Eq.\ \eqref{eq:GK-stretched} holds even if we change the number of particles $N$ (Fig.\ \ref{fig:D-GK}(b)).

Substituting the critical scaling of $D$, $\langle v_y^2\rangle$, and $\tau_\ast$ into the reduced GK formula (Eq.\ \eqref{eq:GK-stretched}),
we derive a relationship between their critical exponents.
In the Newtonian regime, we found the critical scaling as
$D \sim |\Delta\phi|^{-\nu}\dot{\gamma}$, $\langle v_y^2\rangle \sim |\Delta\phi|^{\psi-2\lambda}\dot{\gamma}^2$, and $\tau_\ast \sim |\Delta\phi|^{\lambda-\epsilon}\dot{\gamma}^{-1}$.
Neglecting the weak dependence of $\alpha$ on the control parameters ($\phi$ and $\dot{\gamma}$),
we substitute the critical scaling into $D \sim \tau_\ast\langle v_y^2\rangle$ (Eq.\ \eqref{eq:GK-stretched})
and find $|\Delta\phi|^{-\nu}\dot{\gamma} \sim |\Delta\phi|^{\psi-\lambda-\epsilon}\dot{\gamma}$.
Equating the exponents of $|\Delta\phi|$ on both sides, we obtain
\begin{equation}
\nu + \psi - \epsilon = \lambda~.
\label{eq:sc_re}
\end{equation}
Because this relation (Eq.\ \eqref{eq:sc_re}) connects the critical exponents for different quantities,
we consider Eq.\ \eqref{eq:sc_re} to be a \emph{scaling relation} (by analogy with critical phenomena).
Notice that the exponents of $\dot{\gamma}$ on both sides of Eq.\ \eqref{eq:GK-stretched} coincide.

On the other hand, if $\phi>\phi_J$ and $\dot{\gamma}t_0\geq 10^{-5}$, the critical scaling is
$D \sim |\Delta\phi|^{0.3\lambda-\nu}\dot{\gamma}^{0.7}$, $\langle v_y^2\rangle \sim |\Delta\phi|^{\psi-\lambda}\dot{\gamma}$, and $\tau_\ast \sim |\Delta\phi|^{0.3\lambda-\epsilon}\dot{\gamma}^{-0.3}$.
Substituting them into Eq.\ \eqref{eq:GK-stretched}, we find $|\Delta\phi|^{0.3\lambda-\nu}\dot{\gamma}^{0.7} \sim |\Delta\phi|^{\psi-0.7\lambda-\epsilon}\dot{\gamma}^{0.7}$.
Equating the exponents of $|\Delta\phi|$ on both sides, we find $\nu + \psi - \epsilon = \lambda$ which is exactly the same with Eq.\ \eqref{eq:sc_re}.
Note that the exponents of $\dot{\gamma}$ on both sides of Eq.\ \eqref{eq:GK-stretched} coincide.
Moreover, $D \sim \dot{\gamma}^{1-\nu/\lambda}$, $\langle v_y^2\rangle \sim \dot{\gamma}^{\psi/\lambda}$, and $\tau_\ast \sim \dot{\gamma}^{-\epsilon/\lambda}$
in the $\Delta\phi$-independent critical regime.
From Eq.\ \eqref{eq:GK-stretched}, we find $\dot{\gamma}^{1-\nu/\lambda} \sim \dot{\gamma}^{(\psi-\epsilon)/\lambda}$
so that the exponents satisfy $\nu + \psi - \epsilon = \lambda$,\ i.e.\ the scaling relation, Eq.\ \eqref{eq:sc_re}.
We stress that our estimates of the critical exponents are in accord with Eq.\ \eqref{eq:sc_re} as $\nu + \psi - \epsilon \simeq 3.76$ and $\lambda \simeq 3.87$.
%

\emph{A long-time tail and divergence of the diffusion coefficient.}---
If the system is in the quasi-static regime above jamming,\ i.e.\ $\phi>\phi_J$ and $\dot{\gamma}t_0\leq 10^{-6}$, the data of $C(\tau)$ cannot be fitted to the stretched exponential function.
Figure \ref{fig:autocorrelation}(b) shows double-logarithmic plots of $C(\tau)$, where $\dot{\gamma}t_0=10^{-6}$ and $\phi$ increases as listed in the legend.
We observe power-law decay, $C(\tau)\sim (\dot{\gamma}\tau)^{-1}$ (dashed line), in the long-time limit,\ i.e.\ a \emph{long-time tail}, if $\phi>\phi_J$.
The data for $\phi=0.90$ are well fitted to a function, $C(\tau)=\left(1+c\dot{\gamma}\tau\right)^{-1}$ with $c\simeq 5.6\times 10^3$ (solid line).
Note that similar power-law decay of velocity auto-correlation function was found at the onset of jamming transition, $\phi\simeq\phi_J$ \cite{diff_shear_md3},
and in MD simulations of granular particles sheared at constant pressure \cite{muI1,muI2}.

If the long-time tail is observed as $C(\tau)\sim(\dot{\gamma}\tau)^{-1}$ (dashed line in Fig.\ \ref{fig:autocorrelation}(b)),
the time integral in the GK formula (Eq.\ (\ref{eq:GK})) diverges so that $D\rightarrow\infty$ \cite{SM}.
In recent numerical studies of sheared amorphous solids,
it was reported that the diffusion coefficient over the shear rate scales as $D/\dot{\gamma}\sim L^\beta$ with the system length $L$ and exponent $\beta\geq 1$
\cite{diff_shear_md7,diff_shear_md8,diff_shear_md9,diff_shear_md10,diff_shear_ep1,diff_shear_ep2,diff_shear_ep0,saitoh14}.
This means that the diffusion coefficient diverges in the thermodynamic limit, $L\rightarrow\infty$, as is consistent with the long-time tail.
However, the finite-size scaling, $D/\dot{\gamma}\sim L^\beta$, cannot be explained by our analysis and is left as a future work.

\emph{Discussion.}---
We have studied shear-induced diffusion of two-dimensional soft athermal particles near jamming by MD simulations.
Many previous works focused on the relation between the shear-induced diffusion coefficient $D$ and length scales
such as the cluster size $\xi$ \cite{diff_shear_md4,diff_shear_md5,diff_shear_md6,saitoh14}
and system length $L$ \cite{diff_shear_md7,diff_shear_md8,diff_shear_md9,diff_shear_md10,diff_shear_ep1,diff_shear_ep2,diff_shear_ep0}.
In contrast, we have made a link between $D$ and the relaxation time $\tau_\ast$ through the reduced GK formula as $D \sim \tau_\ast\langle v_y^2\rangle$.
The critical scaling of the mean squared transverse velocity $\langle v_y^2\rangle$
is consistent with the viscosity divergence below jamming and the scaling of yield stress above jamming \cite{rheol0,pdf1,rheol11,rheol15,rl0,saitoh15}.
Except for quasi-static flows above jamming,
we have numerically confirmed that the non-trivial dependence of $\tau_\ast$ on the control parameters, $\phi$ and $\dot{\gamma}$, can be described by the critical scaling.
If the system is in the quasi-static regime above jamming, the velocity auto-correlation function $C(\tau)$ exhibits a long-time tail.
In this case, the time integral in the GK relation diverges, implying finite-size effects on the diffusion coefficient
\cite{diff_shear_md7,diff_shear_md8,diff_shear_md9,diff_shear_md10,diff_shear_ep1,diff_shear_ep2,diff_shear_ep0}.
Recently, it was reported that the shear-induced diffusion coefficient of three-dimensional particles does not diverge in the thermodynamic limit \cite{diff_shear_md10}.
This means that the same scenario in classical fluids, where the GK relation is broken by a long-time tail in two dimensions but is \emph{not} in three dimensions \cite{Hansen},
can be applied to non-equilibrium athermal systems.
Nevertheless, we need to examine whether $C(\tau)$ exhibits a long-time tail in three dimensions
and also have to discuss the relation to the long-time tail observed in granular flows \cite{ltail0,ltail1,ltail2,ltail3,ltail4}.
Moreover, we have proposed a scaling relation between the critical exponents, where our numerical results are in accord with it.
In future, it is necessary to investigate theoretical background to these critical exponents.
Note that our system sizes (typically $N=8192$) are not sufficiently large and the study of much larger system sizes is left as a future work.
%
\begin{acknowledgments}
We thank B. P. Tighe, L. Berthier, H. Hayakawa, M. Otsuki, and S. Takada for fruitful discussions.
This work was supported by KAKENHI Grant Nos.\
20H01868, 21H01006, 22K03459,
JPMJFR212T, 20H05157, 20H00128, 19K03767, 18H01188 from JSPS.
T.K. acknowledges support by the JST FOREST Program (grant no. JPMJFR212T), AMED Moonshot Program (grant no. JP22zf0127009), JSPS KAKENHI (grant no. JP24H02203), and Takeda Science Foundation.
\end{acknowledgments}
\appendix
\section{Critical scaling}
\label{sec:critical}
In this appendix, we explain technical details of our critical scaling.
First, we introduce a scaling function for an observable and explain our method for estimating critical exponents (Sec.\ \ref{sub:exponent}).
Next, we demonstrate scaling data collapses of the diffusivity over the shear rate, $D/\dot{\gamma}$, and $x$-component of mean squared velocity fluctuation, $\langle\delta v_x^2\rangle$ (Sec.\ \ref{sub:collapse}).
We also examine our estimates of the critical exponents by comparing our results with the previous studies \cite{diff_shear_md0,rheol17} (Sec.\ \ref{sub:compare}).
\subsection{Scaling functions and critical exponents}
\label{sub:exponent}
We write a physical quantity (observable) dependent on both the shear rate $\dot{\gamma}$ and proximity to jamming, $\Delta\phi\equiv\phi-\phi_c$, as $A\left(\dot{\gamma},\Delta\phi\right)$.
We assume a \emph{scaling ansatz} for the observable as
\begin{equation}
A\left(\dot{\gamma},\Delta\phi\right) = \left|\Delta\phi\right|^\alpha f_\pm\left(\frac{\dot{\gamma}}{\left|\Delta\phi\right|^\beta}\right)~,
\label{eq:scaling_A}
\end{equation}
where $f_+(x)$ and $f_-(x)$ denote \emph{scaling functions} above and below the jamming transition, respectively.
On the right-hand-side of Eq.\ \eqref{eq:scaling_A}, $\alpha$ and $\beta$ are introduced as \emph{critical exponents}.
Let us rewrite Eq.\ \eqref{eq:scaling_A} as
\begin{eqnarray}
A\left(\dot{\gamma},\Delta\phi\right)
&=& \dot{\gamma}^\frac{\alpha}{\beta}\left(\frac{\left|\Delta\phi\right|^\beta}{\dot{\gamma}}\right)^\frac{\alpha}{\beta}f_\pm\left(\frac{\dot{\gamma}}{\left|\Delta\phi\right|^\beta}\right) \nonumber\\
&\equiv& \dot{\gamma}^\frac{\alpha}{\beta} g_\pm\left(\frac{\left|\Delta\phi\right|^\beta}{\dot{\gamma}}\right)~,
\label{eq:scaling_func_g}
\end{eqnarray}
where we defined new scaling functions as
\[
g_\pm(x) \equiv x^\frac{\alpha}{\beta} f_\pm\left(\frac{1}{x}\right)~.
\]
We assume that $g_\pm\left(\left|\Delta\phi\right|^\beta/\dot{\gamma}\right)$ can be rewritten as \cite{rheol18}
\begin{eqnarray}
g_\pm\left(\frac{\left|\Delta\phi\right|^\beta}{\dot{\gamma}}\right)
&=& g_\pm\left(\left\{\frac{\left|\Delta\phi\right|}{\dot{\gamma}^\frac{1}{\beta}}\right\}^\beta\right) \nonumber\\
&\equiv& F\left(\frac{\Delta\phi}{\dot{\gamma}^\frac{1}{\beta}}\right)~.
\label{eq:scaling_func_F}
\end{eqnarray}
Here, the function $F(x)$ is continuous around the jamming transition, $\Delta\phi=0$,
such that it depends on $\Delta\phi$ but not on the absolute value, $\left|\Delta\phi\right|$.
Substituting Eq.\ \eqref{eq:scaling_func_F} into Eq.\ \eqref{eq:scaling_func_g}, we find
\begin{eqnarray}
A\left(\dot{\gamma},\Delta\phi\right) = \dot{\gamma}^\frac{\alpha}{\beta} F\left(\frac{\Delta\phi}{\dot{\gamma}^\frac{1}{\beta}}\right)~.
\label{eq:scaling_A_F}
\end{eqnarray}

To determine the unknown function $F(x)$, we expand it into a polynomial as
\begin{equation}
F(x) = a_0 + a_1 x + a_2 x^2 + \cdots + a_5 x^5~,
\label{eq:scaling_F_poly}
\end{equation}
where $a_i$ ($i=0,1,\cdots,5$) represents the coefficient.
Then, we estimate not only the critical exponents, $\alpha$ and $\beta$, in Eq.\ \eqref{eq:scaling_A_F},
but also the coefficients $a_i$ in Eq.\ \eqref{eq:scaling_F_poly} from the numerical data of $A\left(\dot{\gamma},\Delta\phi\right)$.
%
\subsection{Scaling data collapses}
\label{sub:collapse}
Introducing a scaling ansatz for the diffusivity over the shear rate as
\begin{equation}
\left|\Delta\phi\right|^\nu \frac{D}{\dot{\gamma}} = h_\pm\left(\frac{\dot{\gamma}}{\left|\Delta\phi\right|^\lambda}\right)~,
\label{eq:scaling_diff}
\end{equation}
we estimate the critical exponents, $\nu$ and $\lambda$.
If we replace $\nu$ and $\lambda$ with $-\alpha$ and $\beta$, respectively, Eq.\ \eqref{eq:scaling_diff} corresponds to Eq.\ \eqref{eq:scaling_A}.
Therefore, from Eq.\ \eqref{eq:scaling_A_F}, the diffusivity over the shear rate is given by
\begin{equation}
\frac{D}{\dot{\gamma}} = \dot{\gamma}^{-\frac{\nu}{\lambda}} F\left(\frac{\Delta\phi}{\dot{\gamma}^\frac{1}{\lambda}}\right)~.
\label{eq:scaling_diff_F}
\end{equation}

\begin{figure}
\includegraphics[width=\columnwidth]{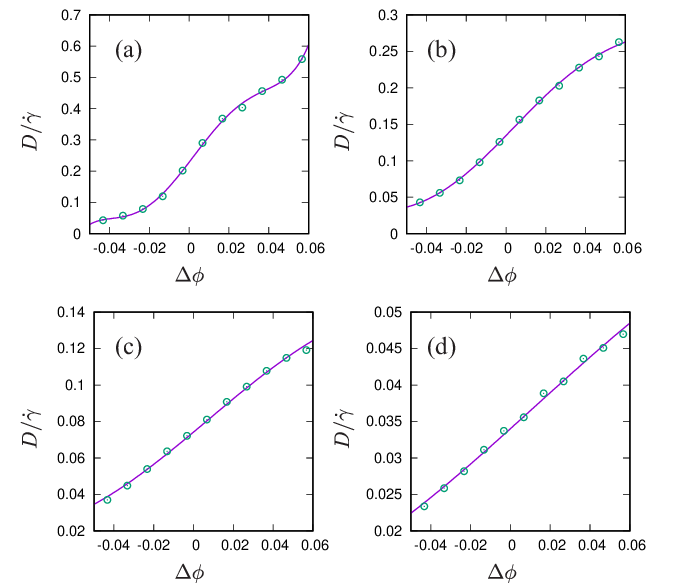}
\caption{
Relations between the diffusivity over the shear rate $D/\dot{\gamma}$ and proximity to jamming $\Delta\phi$,
where the scaled shear rate increases as (a) $\dot{\gamma}t_0=10^{-5}$, (b) $10^{-4}$, (c) $10^{-3}$, and (d) $10^{-2}$.
The circles are numerical results and the solid lines represent the right-hand-side of Eq.\ \eqref{eq:scaling_diff_F}.
\label{fig:demo}}
\end{figure}

We expand $F(x)$ as Eq.\ \eqref{eq:scaling_F_poly} and determine the coefficients, $a_i$ ($i=0,1,\cdots,5$), from the numerical data of $D/\dot{\gamma}$.
Figure \ref{fig:demo} displays $D/\dot{\gamma}$ as a function of $\Delta\phi$.
First, we estimate $a_i$ by fitting the right-hand-side of Eq.\ \eqref{eq:scaling_diff_F} to the data of $D/\dot{\gamma}$ in (a), where the scaled shear rate is given by $\dot{\gamma}t_0=10^{-5}$.
When we fit the right-hand-side of Eq.\ \eqref{eq:scaling_diff_F} to the data, we fix the scaling exponents to $\nu=1$ and $\lambda=4$, which are taken from our previous work \cite{saitoh14}.
Then, we determined the coefficients as
\begin{eqnarray}
a_0 &\simeq&  1.29 \times 10^{-2}~,\nonumber\\
a_1 &\simeq&  2.63 \times 10^{-2}~,\nonumber\\
a_2 &\simeq&  3.59 \times 10^{-3}~,\nonumber\\
a_3 &\simeq& -2.87 \times 10^{-2}~,\nonumber\\
a_4 &\simeq& -1.10 \times 10^{-3}~,\nonumber\\
a_5 &\simeq&  1.82 \times 10^{-2}~.\nonumber
\end{eqnarray}
Next, we fit the right-hand-side of Eq.\ \eqref{eq:scaling_diff_F} to the data in (b)-(d),
where the coefficients $a_i$ are fixed and the critical exponents, $\nu$ and $\lambda$, are used for fitting parameters.
In Fig.\ \ref{fig:demo}, the scaled shear rate increases as (b) $\dot{\gamma}t_0=10^{-4}$, (c) $10^{-3}$, and (d) $10^{-2}$,
where the critical exponents are estimated as (b) $(\nu,\lambda) \simeq (1.01,3.97)$, (c) $(0.976,3.85)$, and (d) $(0.771,3.65)$, respectively.
We take the averages of $\nu$ and $\lambda$ over those used in Figs.\ \ref{fig:demo}(a)-(d) and find mean values as $\nu\simeq 0.939$ and $\lambda\simeq 3.87$.
%
\begin{figure}
\includegraphics[width=\columnwidth]{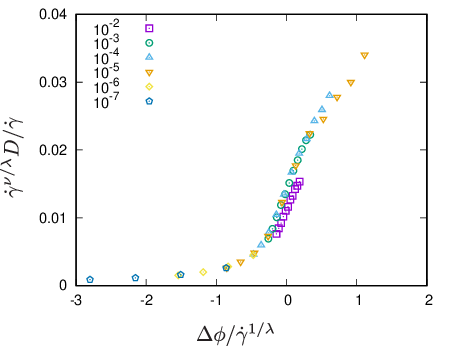}
\caption{
A scaling data collapse of the diffusivity over the shear rate, $D/\dot{\gamma}$,
where the scaling exponents are given by mean values, $\nu\simeq 0.939$ and $\lambda\simeq 3.87$.
The scaled shear rate decreases from $\dot{\gamma}t_0=10^{-2}$ to $10^{-7}$ as listed in the legend.
\label{fig:demo_all}}
\end{figure}
\begin{figure}
\includegraphics[width=\columnwidth]{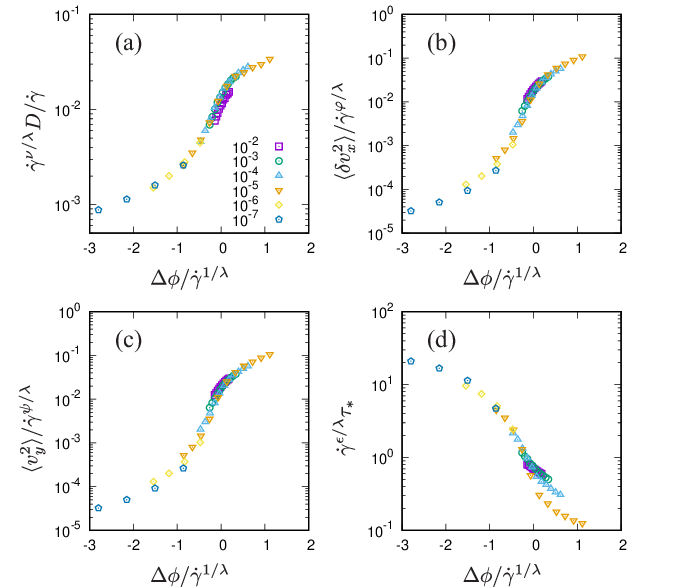}
\caption{
Scaling data collapses of (a) the diffusivity over the shear rate $D/\dot{\gamma}$,
(b) $x$-component of mean squared velocity fluctuation $\langle\delta v_x^2\rangle$,
(c) mean squared transverse velocity $\langle v_y^2\rangle$, and (d) relaxation time $\tau_\ast$,
where the scaled shear rate decreases from $\dot{\gamma}t_0=10^{-2}$ to $10^{-7}$ as listed in the legend of (a).
See the text for the values of critical exponents, $\lambda$, $\nu$, $\varphi$, $\psi$, and $\epsilon$.
\label{fig:demo_log}}
\end{figure}

\begin{figure}
\includegraphics[width=\columnwidth]{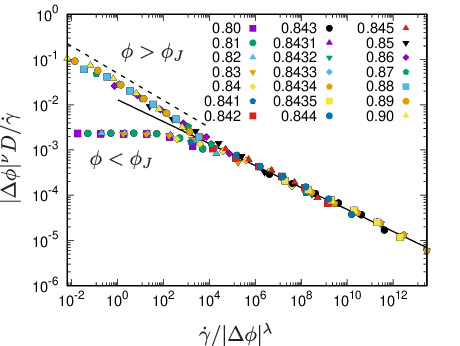}
\caption{
Double logarithmic plots of the diffusivity over the shear rate $D/\dot{\gamma}$ and $\dot{\gamma}$,
where the critical exponents, $\nu\simeq 0.939$ and $\lambda\simeq 3.87$, are used for the scaling data collapse.
The packing fraction $\phi$ increases as listed in the legend.
Note that the data of the largest shear rate ($\dot{\gamma}t_0 = 10^{-1}$) is excluded.
\label{fig:diff}}
\end{figure}

We demonstrate a scaling data collapse of the diffusivity over the shear rate,  $D/\dot{\gamma}$.
Figure \ref{fig:demo_all} shows $D/\dot{\gamma}$ as a function of the proximity to jamming, $\Delta\phi$.
The scaled shear rate decreases as listed in the legend and the critical exponents, $\nu\simeq 0.939$ and $\lambda\simeq 3.87$, are used for the scaling data collapse.
Though the data set with the high shear rate ($\dot{\gamma}t_0=10^{-2}$) slightly deviates from the other data, we can confirm that the scaling data collapse works well.
In addition, Fig.\ \ref{fig:diff} displays $D/\dot{\gamma}$ as a function of $\dot{\gamma}$, where we can see that all the data are well collapsed (see the main text for the meanings of dashed and solid lines).

\begin{figure}
\includegraphics[width=\columnwidth]{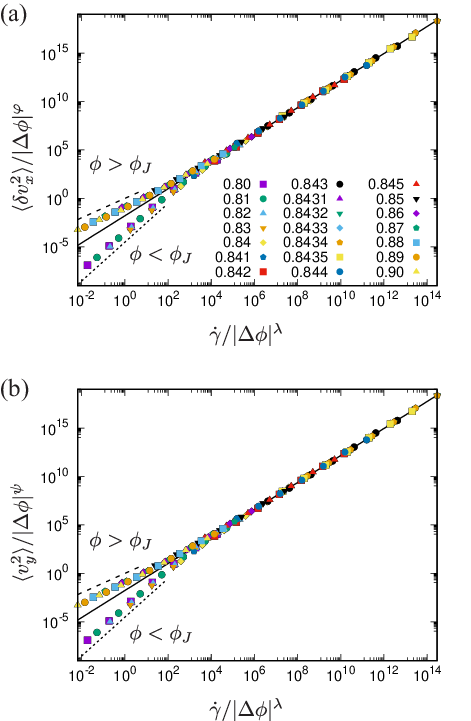}
\caption{
Scaling data collapses of (a) the $x$-component of mean squared velocity fluctuation $\langle\delta v_x^2\rangle$ and (b) mean squared transverse velocity $\langle v_y^2\rangle$,
where the scaling exponents are given by $\varphi\simeq 5.40$, $\psi\simeq 5.39$, and $\lambda=3.87$.
The packing fraction $\phi$ increases as listed in the legend of (a).
\label{fig:dvx}}
\end{figure}

We also demonstrate scaling data collapses of the $x$-component of mean squared velocity fluctuation $\langle\delta v_x^2\rangle$,
mean squared transverse velocity $\langle v_y^2\rangle$, and relaxation time $\tau_\ast$.
Figure \ref{fig:demo_log}(b) shows a scaling data collapse of $\langle\delta v_x^2\rangle$,
where we followed the same procedure to find the critical exponent as $\varphi\simeq 5.40$ (while we fixed $\lambda=3.87$).
As can be seen, all the data with different values of $\dot{\gamma}$ (symbols) are nicely collapsed.
Moreover, we show scaling data collapses of (c) $\langle v_y^2\rangle$ and (d) $\tau_\ast$,
where the critical exponents are estimated by the same procedure as $\psi\simeq 5.39$ and $\epsilon\simeq 2.57$ (while $\lambda=3.87$).
Furthermore, Fig.\ \ref{fig:dvx} displays (a) $\langle\delta v_x^2\rangle$ and (b) $\langle v_y^2\rangle$ as functions of $\dot{\gamma}$,
where the packing fraction $\phi$ increases as listed in the legend of (a).
As can be seen, all the data are well collapsed (see the main text for the meanings of dotted, dashed, and solid lines).

\begin{figure}
\includegraphics[width=\columnwidth]{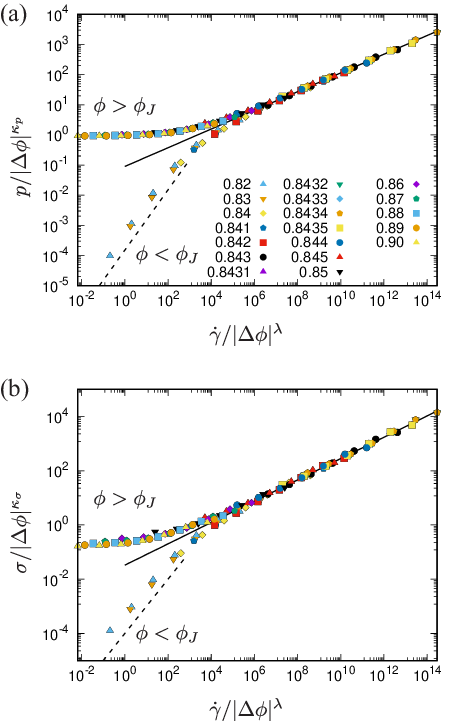}
\caption{
Scaling data collapses of (a) the pressure $p$ and (b) shear stress $\sigma$,
where the scaling exponents are given by $\kappa_p\simeq 1.20$, $\kappa_\sigma\simeq 1.52$, and $\lambda=3.87$.
The packing fraction $\phi$ increases as listed in the legend of (a).
\label{fig:flow}}
\end{figure}

The critical exponent, $\lambda=3.87$, is common not only to $D/\dot{\gamma}$, $\langle\delta v_x^2\rangle$, $\langle v_y^2\rangle$, and $\tau_\ast$ (Fig.\ \ref{fig:demo_log}),
but also to the pressure $p$ and shear stress $\sigma$.
Figure \ref{fig:flow} shows (a) $p$ and (b) $\sigma$ as functions of the shear rate $\dot{\gamma}$, where we assume their scaling ansatz as
\begin{eqnarray}
p &=& \left|\Delta\phi\right|^{\kappa_p} u_\pm\left(\frac{\dot{\gamma}}{\left|\Delta\phi\right|^\lambda}\right)~, \label{eq:scaling_pss} \\
\sigma &=& \left|\Delta\phi\right|^{\kappa_\sigma} w_\pm\left(\frac{\dot{\gamma}}{\left|\Delta\phi\right|^\lambda}\right)~, \label{eq:scaling_sss}
\end{eqnarray}
respectively.
On the right-hand-sides of Eqs.\ \eqref{eq:scaling_pss} and \eqref{eq:scaling_sss},
the scaling functions are given by $u_+(x)\sim\mathrm{const.}$ and $w_+(x)\sim\mathrm{const.}$ above jamming ($\phi>\phi_J$), and $u_-(x)\sim x$ and $w_-(x)\sim x$ below jamming ($\phi<\phi_J$), if $x$ is small.
The scaling exponents are estimated as $\kappa_p\simeq 1.20$ and $\kappa_\sigma\simeq 1.52$, and we confirm that all the data of $p$ and $\sigma$ are nicely collapsed (Fig.\ \ref{fig:flow}).
The dashed lines in Fig.\ \ref{fig:flow} indicate the linear scaling, $u_-(x)\sim x$ and $w_-(x)\sim x$.
On the other hand, the $\Delta\phi$-independent critical regimes are represented by the solid lines,
where their slopes are given by (a) $\kappa_p/\lambda$ and (b) $\kappa_\sigma/\lambda$.
%
\subsection{Comparison with previous studies}
\label{sub:compare}
We examine our estimate of the critical exponents for the diffusivity over the shear rate.
In Ref.\ \cite{diff_shear_md0}, the shear-induced diffusion coefficient is described as
\begin{equation}
\frac{D}{\dot{\gamma}^{q_D}} = F\left(\frac{\Delta\phi}{\dot{\gamma}^{1/(\beta+\Delta)}}\right)
\label{eq:olsson}
\end{equation}
with a scaling function $F(x)$ (see Fig.\ 2 in Ref.\ \cite{diff_shear_md0}).
The critical exponents were reported as $q_D=0.78$ and $1/(\beta+\Delta)=0.275$.
If we rewrite Eq.\ \eqref{eq:scaling_diff_F} as
\begin{equation}
\frac{D}{\dot{\gamma}^{1-\frac{\nu}{\lambda}}} = F\left(\frac{\Delta\phi}{\dot{\gamma}^\frac{1}{\lambda}}\right)~,
\label{eq:scaling_diff_F_2}
\end{equation}
we can compare $1-\nu/\lambda$ and $1/\lambda$ with $q_D$ and $1/(\beta+\Delta)$, respectively.
Substituting our estimates, $\nu\simeq 0.939$ and $\lambda\simeq 3.87$,
we find that $1-\nu/\lambda\simeq 0.757$ and $1/\lambda\simeq 0.258$ which well agree with the values of $q_D$ and $1/(\beta+\Delta)$, respectively.

\begin{figure}
\includegraphics[width=\columnwidth]{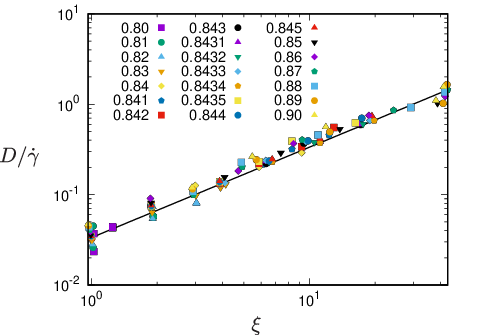}
\caption{
Scatter plots of the diffusivity over the shear rate $D/\dot{\gamma}$ and correlation length $\xi$,
where the packing fraction $\phi$ increases as listed in the legend.
The solid line represents the proportionality, $D/\dot{\gamma}\propto\xi$.
\label{fig:diff_corl}}
\end{figure}

In Ref.\ \cite{rheol17}, the correlation length is described as
\begin{equation}
\xi = \dot{\gamma}^{-1/z}G\left(\frac{\Delta\phi}{\dot{\gamma}^{1/z\nu}}\right)
\label{eq:olsson_xi}
\end{equation}
with a scaling function $G(x)$ (see Fig.\ 7 in Ref.\ \cite{rheol17}).
The critical exponents are given by $1/z=0.27$ and $1/z\nu=0.26$.
As shown in Fig.\ \ref{fig:diff_corl}, the diffusion coefficient over the shear rate is proportional to the length scale as $D/\dot{\gamma}\propto\xi$
and thus we can directly compare Eq.\ \eqref{eq:scaling_diff_F} with Eq.\ \eqref{eq:olsson_xi}.
Substituting our estimates, $\nu\simeq 0.939$ and $\lambda\simeq 3.87$, into Eq.\ \eqref{eq:scaling_diff_F},
we find $\nu/\lambda\simeq 0.243$ and $1/\lambda\simeq 0.258$, which are close to the values of $1/z$ and $1/z\nu$, respectively.
Therefore, our results of the critical exponents for the shear-induced diffusion coefficient are consistent with the previous works \cite{diff_shear_md0,rheol17}.
%
\section{The Green-Kubo relation and long-time tail}
\label{sec:stretch}
In this appendix, we provide some details of the Green-Kubo (GK) relation and long-time tail of velocity auto-correlation function.
First, we introduce the GK relation for the diffusivity and derive the reduced GK formula (Sec.\ \ref{sub:red-GK}).
Then, we show that the velocity auto-correlation function exhibits a long-time tail if the system is slowly sheared above jamming (Sec.\ \ref{sub:longtimetail}).
\subsection{The reduced Green-Kubo formula}
\label{sub:red-GK}
The GK relation for diffusion coefficient $D$ is given by
\begin{equation}
D = \left\langle v_y^2 \right\rangle \int_0^\infty C(\tau) d\tau~,
\label{eq:GK_diffusion}
\end{equation}
where $\left\langle v_y^2 \right\rangle$ is the $y$-component of mean squared velocity and $C(\tau)$ is the normalized velocity auto-correlation function.
If the system is \emph{not} in a quasi-static regime above jamming (see Sec.\ \ref{sub:longtimetail}),
our numerical results of $C(\tau)$ are well described by a stretched exponential function as
\begin{equation}
C(\tau) = \exp\left[-\left(\frac{\tau}{\tau_\ast}\right)^\alpha\right]~,
\label{eq:stretched_exp}
\end{equation}
where $\alpha$ is the stretching exponent and $\tau_\ast$ is defined as a relaxation time.
Integrating Eq.\ \eqref{eq:stretched_exp} over the time interval from $\tau=0$ to $\infty$, we find
\begin{equation}
\int_0^\infty C(\tau) d\tau = \int_0^\infty \exp\left[-\left(\frac{\tau}{\tau_\ast}\right)^\alpha\right] d\tau~.
\label{eq:integral_1}
\end{equation}
If we introduce a new variable as $t\equiv\left(\tau/\tau_\ast\right)^\alpha$, we find
\begin{eqnarray}
dt &=& \frac{\alpha}{\tau_\ast}\left(\frac{\tau}{\tau_\ast}\right)^{\alpha-1} d\tau \nonumber\\
&=& \frac{\alpha}{\tau_\ast} t^{1-1/\alpha} d\tau \nonumber~,
\end{eqnarray}
where we used $\tau/\tau_\ast = t^{1/\alpha}$.
Thus, substituting
\[
d\tau = \frac{\tau_\ast}{\alpha} t^{1/\alpha-1} dt
\]
into Eq.\ \eqref{eq:integral_1}, we find
\begin{eqnarray}
\int_0^\infty C(\tau) d\tau &=& \frac{\tau_\ast}{\alpha} \int_0^\infty e^{-t} t^{1/\alpha-1} dt \nonumber\\
&=& \frac{\tau_\ast}{\alpha} \Gamma\left(\frac{1}{\alpha}\right)~,
\label{eq:integral_2}
\end{eqnarray}
where
\[
\Gamma(x) \equiv \int_0^\infty e^{-t} t^{x-1} dt \hspace{5mm} (x>0)
\]
is the gamma function.
Therefore, if the auto-correlation function is described by the stretched exponential function,
the diffusion coefficient,\ Eq.\ \eqref{eq:GK_diffusion}, is given by
\begin{equation}
D = \frac{\tau_\ast \left\langle v_y^2 \right\rangle}{\alpha} \Gamma\left(\frac{1}{\alpha}\right)~.
\label{eq:GK_diffusion_gamma}
\end{equation}

\begin{figure}
\includegraphics[width=\columnwidth]{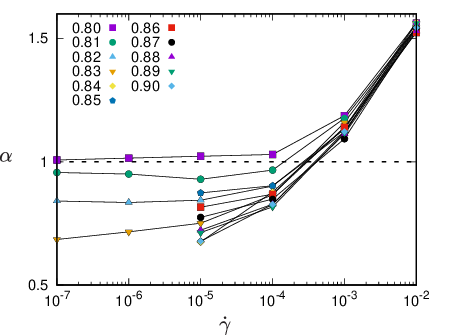}
\caption{
The stretching exponent $\alpha$ as a function of the shear rate $\dot{\gamma}$, where the packing fraction $\phi$ increases as listed in the legend.
\label{fig:exp}}
\end{figure}

Figure \ref{fig:exp} displays the stretching exponent $\alpha$ as a function of the shear rate $\dot{\gamma}$.
If the packing fraction is low enough,\ e.g. $\phi=0.80$,
the stretching exponent converges to unity in the quasi-static limit,\ i.e.\ $\alpha\rightarrow 1$ if $\dot{\gamma}\rightarrow 0$.
This means that the normalized velocity auto-correlation function exponentially decays to zero, $C(\tau)=e^{-\tau/\tau_\ast}$, if the system is in the Newtonian regime.
%
\subsection{Long-time tail}
\label{sub:longtimetail}
\begin{figure}
\includegraphics[width=\columnwidth]{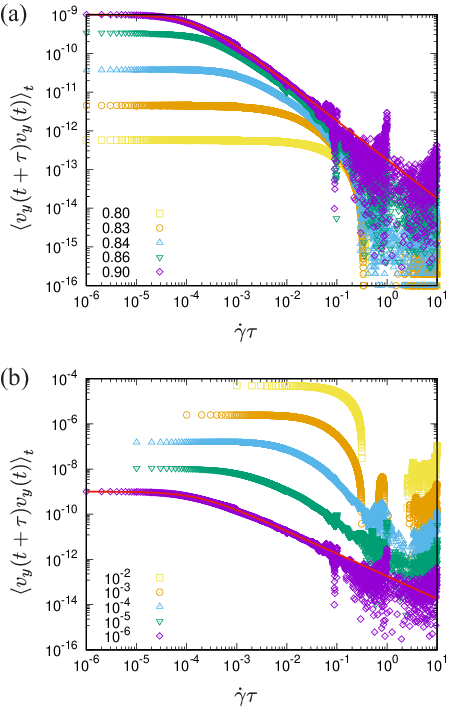}
\caption{
Double logarithmic plots of the velocity auto-correlation function, $\langle v_y(t+\tau)v_y(t)\rangle_t$.
(a) The packing fraction $\phi$ increases as listed in the legend, where the scaled shear rate is given by $\dot{\gamma}t_0 = 10^{-6}$.
(b) The scaled shear rate $\dot{\gamma}t_0$ decreases as listed in the legend, where $\phi=0.90$.
The solid lines in (a) and (b) represent the function,\ Eq.\ \eqref{eq:longtimetail}, fitted to the data for $\phi=0.90$ and $\dot{\gamma}t_0 = 10^{-6}$.
\label{fig:longtimetail}}
\end{figure}

If the system is in a quasi-static regime above jamming,\ i.e.\ if $\dot{\gamma}t_0\ll 1$ and $\phi>\phi_J$,
the velocity auto-correlation function, $\langle v_y(t+\tau)v_y(t)\rangle_t$, exhibits a \emph{long-time tail}.
Figure \ref{fig:longtimetail} displays our numerical results of the velocity auto-correlation function.
In Fig.\ \ref{fig:longtimetail}(a), the packing fraction $\phi$ increases across the jamming transition density $\phi_J$,
where the shear rate is given by $\dot{\gamma}t_0 = 10^{-6}$.
As can be seen, the velocity auto-correlation function exhibits the power-law decay (long-time tail) if $\phi>\phi_J\simeq 0.8433$.
The solid line represents a function,
\begin{equation}
\langle v_y(t+\tau)v_y(t)\rangle_t = \frac{\langle v_y^2\rangle}{1+c\dot{\gamma}\tau}~,
\label{eq:longtimetail}
\end{equation}
fitted to the data for $\phi=0.90$, where $c\simeq 5.6\times 10^3$ is a fitting parameter.
In Fig.\ \ref{fig:longtimetail}(b), the packing fraction is fixed to $\phi=0.90>\phi_J$, where the scaled shear rate decreases as listed in the legend.
One can see that the power-law behavior of the velocity auto-correlation function is more enhanced with the decrease of the shear rate.

If we substitute Eq.\ \eqref{eq:longtimetail} to Eq.\ \eqref{eq:GK_diffusion}, the diffusion coefficient is given by
\begin{eqnarray}
D &=& \left\langle v_y^2 \right\rangle \int_0^\infty \frac{d\tau}{1+c\dot{\gamma}\tau} \nonumber\\
&=& \frac{\left\langle v_y^2 \right\rangle}{c\dot{\gamma}} \left[\log\left(1+c\dot{\gamma}\tau\right)\right]_0^\infty \nonumber\\
&=& \frac{\left\langle v_y^2 \right\rangle}{c\dot{\gamma}} \log(\infty)~.
\label{eq:diverge_D}
\end{eqnarray}
Therefore, $D$ diverges since $\left\langle v_y^2 \right\rangle$ is finite in $\dot{\gamma}t_0\ll 1$ and $\phi>\phi_J$.
%
\bibliography{SE_relation}
\end{document}